# The Atomic Hydrogen Cloud in the Saturnian System

W.-L. Tseng[1], R. E. Johnson[2] and W.-H. Ip[3]

1. Division of Space Science and Engineering, Southwest Research Institute, San Antonio TX, USA

2. Department of Materials Science and Engineering, University of Virginia, Charlottesville VA, USA

3. Institute of Astronomy, National Central University, Chungli City, Taoyuan County, Taiwan

## Abstract

The Voyager flyby observations revealed that a very broad doughnut shaped distribution of the hydrogen atoms existed in the Saturnian magnetosphere (Broadfoot et al., 1981). The origin of the atomic hydrogen cloud has been debated ever since. Shemansky and Hall (1992) showed that this cloud had an azimuthal asymmetry dependent on local time with a higher intensity on the dusk side. Smyth and Marconi (1993) suggested that the cumulative effect of solar radiation pressure on the long-term orbital motion of the hydrogen atoms escaping from Titan could explain the observed morphology. Ip (1996) showed that, in addition to the Titan's hydrogen torus, the sun-lit hemisphere of Saturn's atmosphere and/or the ring system could be major sources of the hydrogen atoms in the inner magnetosphere ($<\sim 10 R_S$). Recent Cassini UVIS observations confirmed the local-time asymmetry but also showed the hydrogen cloud density increases with decreasing distance to Saturn's upper atmosphere with a peak at Saturn. Shemansky et al. (2009) attributed this to hydrogen atoms flowing outward from the Saturn's sun-lit hemisphere due to electron-impact dissociation of $H_2$.

The Saturnian system is also immersed in a vast gas cloud of $H_2O$, $O_2$ and $H_2$ and their dissociation products (OH, O, and H) which originate from the Enceladus plumes, the rings, the inner icy satellites and Titan's $H_2$ torus (e.g., Tseng et al., 2011). Since these neutrals are important sources of $H^+$ for the magnetosphere, we have carried out a global investigation of the atomic hydrogen cloud taking into account all possible sources: 1) Saturn's atmosphere, 2) the $H_2$ atmosphere of main rings, 3) Enceladus' $H_2O$ and OH torus, 4) Titan's $H_2$ torus and 5) the atomic hydrogen directly escaping from Titan. We show that the H ejection velocity and angle distribution are modified by collisions of the hot hydrogen, produced by electron-impact dissociation of $H_2$, with the ambient atmospheric $H_2$ and H. This in turn affects the morphology of the escaping hydrogen as does the morphology of the ionospheric electron distribution. That Saturn's atmosphere is an important source is suggested by the fact that the H cloud peaks well below the ring plane, a feature that, so far, we can not reproduce by the dissociation of the ring $H_2$ atmosphere or other proposed sources. Our simulations show that H directly escaping from Titan is a major contribution in the outer magnetosphere ($>\sim 10 R_S$). The morphology of Titan's H torus, shaped by the solar radiation pressure and Saturn's oblateness, can account for the local time asymmetry near Titan's orbit. Dissociation of $H_2O$ and OH in the Enceladus torus contributes inside $\sim 5\ R_S$, but dissociation of Titan's $H_2$ torus does not due to the significant energy released. The total number of H observed by Cassini UVIS from $>1\ R_S$ to $5\ R_S$ is $\sim 5.7 \times 10^{34}$; our modeling results suggest ~20% from dissociation in the Enceladus torus, ~5-10% from dissociation of ring $H_2$ atmosphere, and ~50% from Titan's H torus implying that ~20% comes from Saturn's upper atmosphere.

## 1. Introduction

The first report of a hydrogen cloud in the Saturnian magnetosphere was given by Broadfoot et al. (1981) after the Voyager flyby. They showed that it had a very broad distribution of hydrogen atoms in a doughnut-shape region between 8-25 $R_S$ and with a vertical thickness of ~14 $R_S$. Since then the source of the atomic hydrogen cloud has been a long standing puzzle. With a more detailed analysis of Voyager UVS data, Shemansky and Hall (1992) found that the atomic hydrogen cloud in the Saturnian magnetosphere exhibited an azimuthal asymmetry that was dependent on local time with a higher intensity on the dusk side. Because the H Ly-α emission intensity increased with decreasing distance to Saturn, they suggested that the sun-lit side of Saturn atmosphere was the primary source of H produced by electron impact dissociation of $H_2$. In addition, a peak in the emission intensity near 20 $R_S$ indicated a significant contribution from Titan.

Smyth and Marconi (1993) suggested that a source from the sunlit Saturn hemisphere could not explain the observed complex 3D morphology. Since the lifetime of the atomic hydrogen in the vicinity of Titan (~$10^8$ sec) is much longer than Titan's orbital period ($1.4 \times 10^6$ s), they showed that the cumulative effect of solar radiation pressure is very important for describing the long-term orbital motion of the hydrogen atoms prior to their loss. The initial circular orbits would be shaped into highly eccentric ones forming a doughnut-shaped distribution. Ip (1996) simulated the morphology of Titan's atomic hydrogen cloud using Monte-Carlo calculations that took into account Saturn's gravity, the J2 term due to its oblateness and the solar radiation pressure. This modeling showed that Titan's atomic hydrogen cloud was asymmetric, as a function of local time, with depletion on the predawn side. Ip (1996) also pointed out that Saturn's sun-lit hemisphere and/or the ring system could be major sources of the hydrogen atoms in the inner magnetosphere ($< \sim 10$ $R_S$). From the recent Cassini UVIS observations, a much broader distribution of the atomic hydrogen cloud extending beyond 40 $R_S$ in the equatorial plane was found by Melin et al. (2009). It also exhibits local-time asymmetry and with the intensity increasing toward the top of Saturn's atmosphere (Shemansky et al., 2009). The density peak observed at SOI appeared on Saturn's sun-lit hemisphere at latitude of -13.5° while the subsolar point was at -23.6°. The evidence shown above appeared to support the theory that the hydrogen atoms were flowing outward from the Saturn's sun-lit hemisphere as a result of the electron-impact dissociation of $H_2$ as proposed by Shemansky et al. (2009). However, there are still some questions left unanswered. For example, although they suggested that atomic hydrogen escapes radially from Saturn in a plume-like structure, they failed to provide a reasonable physical model.

In the H Ly-α intensity map observed by Cassini UVIS at SOI, the atomic hydrogen cloud appeared to concentrate around the main ring region (see Fig. 4f in this paper). Our previous work showed that the ring system is an important source of $H_2$ and $O_2$ which are injected into the Saturnian magnetosphere via scattering processes (e.g. Johnson et al., 2006; Tseng et al., 2010; 2011; 2012). Hence, it is reasonable to ask whether the main rings could also be a significant contributor as suggested earlier by Carlson (1980) and others. In addition, the Saturnian system is immersed in a vast neutral gas cloud of $H_2O$, $O_2$, $H_2$ and their dissociative products like OH, O and H. Most of the gas molecules originate from Enceladus' plumes plus small contributions from other inner icy satellites (e.g. Waite et al., 2006). Titan's exosphere is another major source of neutral $H_2$ and H as suggested earlier, and, possibly, $CH_4$ and $N_2$ (e.g. Tucker and Johnson, 2010). Since the neutral hydrogen cloud is an important source of $H^+$ for the magnetosphere, in this paper we carry out a comprehensive investigation of the atomic hydrogen cloud in the Saturnian system. We take into account all possible sources: 1) Saturn's exosphere, 2) the $H_2$ atmosphere of main rings, 3) Enceladus' $H_2O$ and OH torus, 4) Titan's $H_2$ torus and 5) the atomic hydrogen directly escaping from Titan. Our numerical modeling is described in Section 2, where we use the plasma densities and temperatures in the Saturnian magnetosphere (Sect. 2.1) to compute the H lifetime (Sect. 2.2) as well as the H production rate from each source (Sect. 2.3). Descriptions of orbital dynamics and the cumulative effect of radiation pressure in the trajectory simulations are given in Sect. 2.4. For the Saturn exospheric source (Sect. 2.5), we calculate the modification of the velocity distribution of hot hydrogen atoms formed from electron-impact dissociation of $H_2$ due to collisions with the ambient atmospheric $H_2$ and H, which is neglected in Shemansky et al. (2009). The location of the principal source region on Saturn is also discussed in Sect. 2.5.2. The simulated density distributions contributing to the hydrogen cloud from each source are then shown in Section 3. This section also includes a comparison with the Cassini UVIS data (Shemansky et al., 2009 and Melin et al., 2009). A summary is given in Section 4.

## 2. Modeling Descriptions

**2.1 Plasma Environment**

A model of the Saturn's magnetospheric plasma environment was built using the latest Cassini CAPS measurements of the ion density and temperature distributions of $H^+$, $H_2^+$ and $W^+$ given in Sittler et al. (2008) and Thomsen et al. (2010). The electron densities and temperatures obtained using Cassini CAPS were summarized by Schippers et al. (2008). The electron energy spectrum is typically divided to two main components: the dense, relatively cold thermal component, with a peak density at ~4-5 $R_S$, and the tenuous relatively hot suprathermal component, which peaks at ~9.0 $R_S$. Since the latitudinal information on either is insufficient, we use the same variation of both populations and equate the total electron density to the total ion density (sum of $H^+$, $H_2^+$ and $W^+$). This plasma model will be used to calculate the H lifetime throughout the magnetosphere, in addition to the H production rates from Enceladus' neutral $H_2O$ and OH clouds and from Titan's $H_2$ torus as described below.

**2.2 H Lifetime**

The H lifetime is constrained not only by the magnetospheric plasma environment (ion/electron density and temperature) but also by the solar photon flux. The net ionization rate is due to photoionization, electron-impact ionization and ion-molecule reactions. The photoionization rate of H at Saturn's orbit is ~$1.4 \times 10^{-9}$ s$^{-1}$ (Huebner et al. 1992). The electron-impact ionization rate, v, is given by v=$n_e$*σ(u)*u, where u(= $(\pi kT/8m_e)^{1/2}$) is the average electron velocity for a given electron temperature given by Schippers et al. (2008), $n_e$ is the electron number density and σ(u) is the energy-dependent electron-impact cross-section which is taken from Kim and Rudd (1994). The ion-molecule reaction rate α is calculated using α=$n_i$*σ($E_i$)*$V_i$ with the relative ion-neutral collision speed $V_i$, the energy-dependent charge exchange cross-section σ($E_i$), and the local ion density $n_i$. Here we consider the charge exchange collisions between neutral H and the protons (H$^+$) as well as the water-group ions (W$^+$) whose energy-dependent cross-sections are taken from Tawara et al. (1985). This chemical network is used to compute the spatial variation of neutral H lifetime in a 2D Saturnian magnetosphere with azimuthally symmetry.

Figure 1 presents the variation with distance from Saturn of the neutral H lifetime at the equator. Over the main rings (which is not shown in Figure 1), photoionization dominates and H has a very long lifetime ~ $7.4 \times 10^8$ sec. In inner region inside of ~9 $R_S$, the minimum lifetime of H is ~$10^7$ sec which is mainly constrained by charge exchange collisions. The H lifetime near 17-20 Rs reaches ~$2 \times 10^8$ sec. Outside Titan's orbit, ~$2 \times 10^8$ sec is assumed by taking into consideration the charge exchange with the solar protons (Marconi and Smyth, 2006).

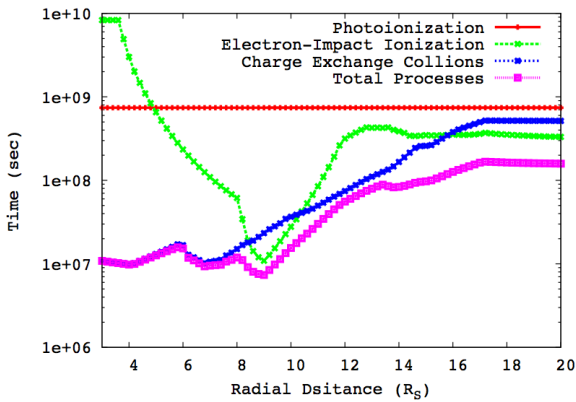

Figure 1: Neutral H lifetime (s) at the equator as a function of distance from Saturn (in unit of Saturn's radius: $R_S$). Lifetimes for photoionization (red), electron-impact ionization (green), charge-exchange (blue) and all processes combined (pink).

**2.3 Estimates of the H source rates**

**2.3.1 Saturn's exosphere**

As described above, based on the analysis of the Voyager UVS and Cassini UVIS data, Shemansky and Hall (1992) and Shemansky et al. (2009) claimed that ignoring the features on Saturn's disc (airglow and the aurora) the broad distribution of the atomic hydrogen cloud observed with local time asymmetry in the Saturnian system is formed from the hydrogen atoms escaping from Saturn as a result of electron-impact dissociation of $H_2$. The source rate of $\sim 3 \times 10^{30}$ H s$^{-1}$ given by Shemansky et al. (2009) was approximated using the total number of H observed from >1 $R_S$ to 5 $R_S$ ($\sim 5.7 \times 10^{34}$) divided by the average lifetime of the H with ballistic trajectories re-impacting Saturn (~5 hrs). Because there are other possible sources, such as the $H_2$ atmosphere of main rings and the Enceladus neutral torus, their estimate is a rough upper limit. Here we use $\sim c(3 \times 10^{30})$ H s$^{-1}$ so that when we include other sources we can use $c$ to scale the exospheric source to fit the observations.

**2.3.2 The $H_2$ atmosphere of the main rings**

Cassini SOI observations confirmed the existence of the $O_2$ atmosphere of the main rings mainly produced by photolytic decomposition of water ice (e.g. Johnson et al., 2006; Waite et al., 2005; Tokar et al., 2005). The same decomposition mechanism also generates $H_2$ from the icy particles in a stoichiometric equilibrium of a ratio of 2:1, $H_2$ to $O_2$. Since the large ring $O_2$ atmosphere observed by Cassini is enhanced by recycling of oxygen on the ring particle surfaces (Tseng et al., 2011; 2012) and some contribution from the Enceladus torus radicals interacting on the A-ring particles (Tseng and Ip, 2011), the ratio is likely smaller. In previously published work (Tseng et al., 2011; 2012), we showed that the $H_2$ density near equator in the main ring region is ~ $2 \times 10^5$ cm$^{-3}$ with a scale height of ~4,000 km. This is consistent with an $H_2$ source rate of ~$1 \times 10^{27}$ s$^{-1}$ near SOI. Elrod et al. (2012) found that large variations in ion density, temperature, and composition occurred between the Voyager 2 flyby and the Cassini CAPS data from 2004 to 2010 in the region of ~2.3 to ~ 3.5 $R_S$. This was described using a one-box ion chemistry model combining the water products from Enceladus and oxygen from the ring atmosphere. Although the Enceladus outgassing rate itself is variable, we showed that the large variation was likely due to the seasonal variation in the ring atmosphere (Tseng et al., 2012). Therefore, the ring $H_2$ atmosphere would be significantly reduced near equinox (down to ~$1\times10^{25}$ $H_2$ $s^{-1}$). Below we present the estimates at SOI for comparison with the Cassini H Ly-α intensities at about the same time period.

Although H and $H_2O$ are directly ejected from the cold icy ring particles, on return they primarily stick, so that the more volatile $H_2$ and $O_2$ accumulate. The morphology and structure of the $H_2$ atmosphere over the main rings has been modeled by Tseng et al. (2010) and normalized to CAPS data from Elrod et al. (2012). The ring $H_2$ atmosphere can produce hydrogen atoms as a result of photodissociation and electron-impact dissociation. The primary photodissociation process forming H has a rate of $6.3\times10^{10}$ $s^{-1}$ with the total excess energy of ~0.4 eV (Huebner et al., 1992). The other dissociation pathway is discarded due to the large excess energy of ~8.2 eV which results in the energetic H fragments escaping from Saturn. The total number of $H_2$ over the main rings is ~$2.2\times10^{35}$ when the $H_2$ source rate is ~$1\times10^{27}$ $s^{-1}$ at SOI. Thus, a H source rate of ~$2\times(1.4\times10^{26})$ $s^{-1}$ from the ring $H_2$ atmosphere is obtained by photodissociation.

The contribution from the electron-impact dissociation of $H_2$ over the main rings is difficult to determine as there is little plasma information due to one single Cassini pass at a relatively high latitude. The single set of measurements gives an electron temperature over the main ring of about 0.5-2 eV (Coates et al., 2005). Based on the $O_2^+$ and $O^+$ ion densities over the main rings published by Tokar et al. (2005) and Elrod et al. (2012), the electron density would be in the order of magnitude of ~5-10 $cm^{-3}$ under an assumption of equating the electron density to the total ion density (at a high altitude ~0.15 $R_S$). The electron-impact dissociation rate of $H_2$ is ~$3.2\times10^{-11}$ $cm^{-3}$ $s^{-1}$ using $T_e = 1$ eV and $n_e = 5$ $cm^{-3}$ (Shemansky et al., 2009). Ignoring the variations in the electron properties (temperature and density) along the field lines we use an approximate source of H produced by electron impact of ~ $2\times(3.5\times10^{25})$ $s^{-1}$. However, the excess energy from electron-impact dissociation of $H_2$ can be large. For the kinetic energy of H produced from electron-impact on $H_2$, Yoon et al. (2008) showed that the electrons at low energy (<16.5 eV) primarily excite the $H_2$ to the b $^3\Sigma_u^+$ state so the dissociated H have a kinetic energy of a few eV due to the steep repulsive potential curve ((e.g. Shemansky et al., 2009). This is the case over the main rings. But, with increasing electron energy (>16.5 eV), the higher excited states of $H_2$ dominate producing lower energy H fragments because the potential curves for the higher excited states are much less repulsive. As a result of a smaller source rate and a larger excess energy, the contribution of electron-impact dissociation of $H_2$ over the main rings is not a significant contributor.

### 2.3.3 Enceladus' $H_2O$ and OH torus

Cassini discovered that Enceladus has active plumes ejecting mainly water molecules and icy grains from its south polar region, which is the dominant source of $H_2O$, OH and O in the Saturn's inner magnetosphere. A much broader distribution of the O and OH has been observed by Cassini UVIS (Melin et al., 2009) than in the early estimate by Johnson et al. (2006). Farmer (2009) and Cassidy and Johnson (2010) showed that, in addition to charge exchange collisions and the energy release from molecular dissociation, the increased spreading was due to neutral-neutral collisions which take place predominantly in the densest part of Enceladus' water cloud.

Atomic H can be produced from $H_2O$ and OH impacted by photons and electrons. Thus, the broad Enceladus $H_2O$ and OH torus could be an important source of H in the Saturnian magnetosphere. Here we use the density distribution of the water cloud simulated by Cassidy and Johnson (2010). Following the analysis in Tseng et al. (2011), we find that $\sim 2.0 \times 10^{34}$ $H_2O$ molecules are present in the steady state produced by an average Enceladus' source rate of $\sim 1 \times 10^{28}$ $H_2O$ $s^{-1}$. The photodissociation rates of $H_2O$ and OH forming H are $\sim 1.5 \times 10^{-7}$ $s^{-1}$ and $\sim 1.4 \times 10^{-7}$ $s^{-1}$ at Saturn's orbit, respectively (Huebner et al., 1992). Therefore, the H production rate from photodissociation of $H_2O$ is $\sim 3.0 \times 10^{27}$ H $s^{-1}$. The density distribution of the OH cloud is very close to the $H_2O$ cloud except a slightly higher value for $H_2O$ near Enceladus' orbit (Cassidy and Johnson, 2010). For simplicity, the total number of OH is assumed to be the same as $H_2O$. The H produced from the OH cloud is at a rate of $\sim 2.8 \times 10^{27}$ $s^{-1}$. Although photodissociation of $H_2O$ and OH is the dominant process throughout the magnetosphere (Smith et al., 2010), this contribution to the total H would be small since the excess energy is so large that most of the H fragments escape from Saturn ($\Delta E = 3.4 eV$ for $H_2O$ and $\Delta E = 2.0 eV$ for OH from Huebner et al., 1992).

To compute the contribution from the electron-impact dissociation of $H_2O$ and OH, we use the electron density and temperature distributions observed by Cassini CAPS (Schippers et al., 2008) as described in Section 2.1. The energy-dependent cross-sections of $H_2O$ collided by electrons is taken from Itikawa and Maston (2005). Also, the cross-sections of OH can be approximated by that of $H_2O$. Even though the peak of $H_2O$ density is located around 4-5 $R_S$ shown in Cassidy and Johnson (2010), we find that the H production by electron impacts is significant only in a small region between 7-10 $R_S$ due to presence of hot electrons. This is also consistent with the findings in Smith et al. (2010). The total H production rate from Enceladus $H_2O$ torus is $5.0 \times 10^{26}$ H $s^{-1}$, which is approximately one order of magnitude smaller than the photodissociation rate. Electron impact dissociation of OH produces H at a similar rate $\sim 5.0 \times 10^{26}$ $s^{-1}$. The kinetic energy of H after electron-impact dissociation is assumed the same as photodissociation.

### 2.3.4 Titan's $H_2$ torus

The $H_2^+$ ions in the Saturnian magnetosphere have been investigated by Cassini CAPS (Thomsen et al. 2010). Tseng et al. (2010) showed that Titan is a major source for neutral $H_2$ in the region outside of $\sim 6.0 R_S$ and the density distribution of $H_2^+$ estimated from our ion source rates roughly agrees with CAPS observations. Like the $H_2$ atmosphere over the main rings, Titan's $H_2$ torus is a potentially important source for atomic H in the Saturn's system. With a $H_2$ escape rate of $1.0 \times 10^{28}$ $s^{-1}$ from Titan, there are $\sim 2.2 \times 10^{35}$ $H_2$ in steady state of Titan's $H_2$ torus (Tseng et al. 2011). Using the photodissociation rate of $6.25 \times 10^{-10}$ $s^{-1}$ discussed above, the H source rate from photodissociation of Titan's $H_2$ torus is $\sim 2 \times (1.4 \times 10^{26})$ $s^{-1}$. Each H formed from photodissociation has an excess energy of 0.2 eV (Huebner et al., 1992).

The electron densities and temperatures in the Saturnian magnetosphere as described above are used to derive the H production rate by electron-impact dissociation in Titan's $H_2$ torus which is $\sim 2 \times (4.5 \times 10^{26})$ $s^{-1}$. While the neutral $H_2$ density peaks near Titan's orbit, this source of H peaks around 8-12 $R_S$ due to density of hot electrons. As discussed before, the same excess energy of photodissociation assumed for electron-impact dissociation of $H_2$ is reasonable since the electron temperature is mostly larger than 20 eV in the H production region.

### 2.3.5 Titan's exosphere

The main composition of Titan's atmosphere is $N_2$ (~95%), $CH_4$ (~4%) and $H_2$ with trace elements of hydrocarbons. Although it is still debated whether or not Titan has significant atmospheric loss rates for heavy species (e.g. Yelle et al. 2008; Cui et al. 2008; DeLaHaye et al. 2007; Strobel 2008; Tucker and Johnson 2009) within a factor of two, there is much closer agreement on the rate of loss of $H_2$ (e.g., Tucker et al. 2012). For example, a hydrogen mass loss rate of ~$1.6 \times 10^{28}$ amu/s is obtained by Cui et al. (2008) derived from Cassini INMS data was used to populate the $H_2$ torus (Tseng et al. 2011).

From the photochemical modeling proposed by DeLa Haye et al. (2007), the suprathermal populations found in Titan's exosphere were consistent with the photodissociation products, and a hot to thermal neutral atomic hydrogen ratio of $5 \times 10^{-5}$ has been inferred by a method based on Liouville theorem. Krasnopolsky (2009) developed a global-mean model of coupled neutral and ion chemistry to describe the Titan's atmosphere and ionosphere. The escape rate of the neutral hydrogen atoms from its exobase was estimated to be $6.0 \times 10^{27}$ $s^{-1}$. Based on Cassini HDAC measurements, an escape rate of H from Titan is inferred to be ~ $1.74 \times 10^{27}$ $s^{-1}$ by Hedelt et al. (2010) at a Maxwellian thermal flux distribution at an exospheric T=150K which are used in our simulations of Titan's H torus. The H production rates of all proposed sources are summarized in Table 1.

Table 1: A summary of the H source rates

| Source Name | H source rate ($s^{-1}$) and source mechanism |
|---|---|
| Saturn's Exosphere | $c(3 \times 10^{30})$ from electron-impact dissociation of $H_2$ (Shemansky et al., 2009) with a scaling factor $c$ |
| $H_2$ atmosphere of main rings | $2.8 \times 10^{26}$ from photodissociation <br> $0.7 \times 10^{26}$ from electron-impact dissociation |
| Enceladus' $H_2O$ and OH torus | $5.8 \times 10^{27}$ from photodissociation <br> $1.0 \times 10^{27}$ from electron-impact dissociation |

| Titan's $H_2$ torus | $2.8 \times 10^{26}$ from photodissociation |
| | $9.0 \times 10^{26}$ from electron-impact dissociation |
| Titan's exosphere | $1.74 \times 10^{27}$ from thermal escape (Hedelt et al., 2010) |

### 2.4 Orbital Dynamics

A Monte Carlo test-particle model like that used earlier (Tseng et al. 2011), was used to describe the atomic hydrogen torus in the Saturn's magnetosphere. The center of coordinate system is fixed at Saturn. In the equation of motion, the Saturn's gravity, the J2 perturbation due to the oblateness and solar radiation pressure acceleration are all taken into account. The details of our computation of solar radiation pressure acceleration are described in Section 2.4.1. The initial position and the initial weighting factor of each test particle are determined by their parent molecular-cloud morphology (e.g. the ring $H_2$ atmosphere, Enceladus' neutral torus or Titan's $H_2$ torus). The test particles are launched with a local Kepler velocity plus an excess energy from dissociation processes. In addition, because the Titan's orbital period is much shorter than the dynamical lifetime of hydrogen atoms, the injection from Titan can be simply regarded as a ring located 20.2 Rs with a continuous distribution of emission points. The hydrogen test particles isotropically escaping from Titan's orbit are launched with the local Kepler velocity of Titan plus a thermal velocity from a Maxwellian flux distribution at an exospheric temperature T=150K. Since the simulation of H from Saturn's exosphere is more complicated, it will be discussed in Section 2.4.2.

In order to calculate the chemical loss rate as discussed in Section 2.2, the weighting factor of each test particle is reduced by a value of $\exp(-\Delta t/\gamma_{Loss})$ in an integration time step ($\Delta t$) where $\gamma_{Loss}$ is the local destruction timescale. The loss to absorption of the main rings is also included accounting for the optical depth at the point of impact. In addition when a test particle hits Saturn (at $1R_S$) or travels outside 60 $R_S$, the trajectory is terminated. We use M (usually ~100,000) test particles from each source in our Monte-Carlo computations. The conversion factor (C) of our numerical calculations to the real hydrogen number density is: $C = Q/(M/\Delta t)$ with the actual hydrogen source rate, Q.

### 2.4.1 Solar Radiation Pressure

Following the numerical approach in Ip (1996), the solar radiation pressure acceleration is 0.75 of the solar gravitational acceleration at Saturn's orbit, and it is related to two angles, $\Phi$, and $\Omega$. $\Phi$ ($=2\pi t/P_S$) is the phase angle of the orbital motion of Saturn around the Sun with $P_S$, the orbital period of Saturn. $\Omega$ is the obliquity of Saturn if the coordinate system is defined such that the y-axis is tangential to the orbital motion at $\Phi=0°$. Two extreme initial conditions, $\Omega = 25°$ with $\Phi=0°$ (SOI phase) and $\Omega = 0°$ with $\Phi=90°$ (equinox phase) are considered in the modeling to examine if there are significant seasonal variations. Figure 2 shows the geometry of the coordinate system co-moving with Saturn as it orbits the Sun. The cumulative radiation pressure is significant only for the atomic H in Saturn's outer magnetosphere.

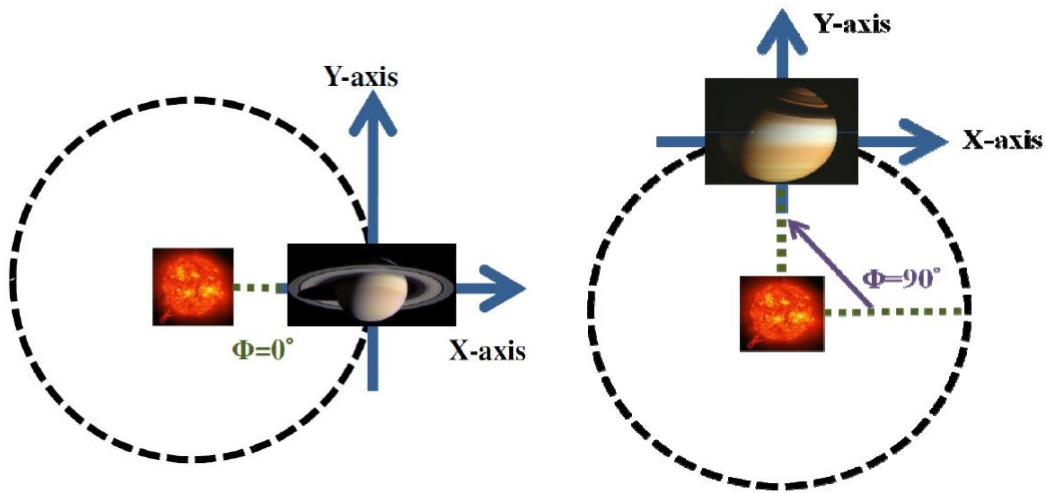

Figure 2: The initial conditions for the geometry of the coordinate system with respect to the solar illumination: (Left) inclination angle $\Omega =25°$ (SOI phase) and the orbital phase angle $\Phi =0°$. (Right) Inclination angle $\Omega =0°$ (equinox phase) and the orbital phase angle $\Phi =90°$.

### 2.5 H escape from Saturn's atmosphere
### 2.5.1 Initial Ejection Velocity Distribution

The atomic H in the Saturn's magnetosphere detected by Voyager UVS and Cassini UVIS (H Ly-α emission) was suggested to be primarily ejected from Saturn's atmosphere as a result of electron-impact dissociation of $H_2$ (Shemansky and Hall, 1992; Shemansky et al.,

2009). The hot H is produced from excitation of the repulsive $H_2$ b state by electrons, as discussed earlier. They also showed that the H fragments could reach energies higher than ~3.0 eV. In addition, the processes must take place within 1-2 scale heights of the exobase (2250 km) to avoid depositing the energy directly into the ambient atmosphere. The scale heights are ~160 km for $H_2$ and ~320 km for H with an atmospheric temperature ~400-500 K (Nagy et al., 2008) near exobase. However, the collisional thermalization of the hot H by the ambient atmospheric $H_2$ and H occurring within 1-2 scale heights can significantly modify the velocity distribution of gas in the putative hydrogen plume of Saturn. Therefore, we have computed a new hydrogen ejection velocity distribution from Saturn taking into account of the collisional cooling effect with the ambient $H_2$ and H. The numerical approach has been presented by Ip (1988) in describing a hot oxygen corona of Mars. We use the H and $H_2$ ambient density profile with variation of altitude in Saturn's atmosphere from Nagy et al. (2009) and Shemansky et al. (2009). The initial velocity distribution of the hot hydrogen produced directly from the electron impact dissociation of $H_2$ is taken from Shemansky et al. (2009) (as shown in Figure 3a) assuming that the reaction of the lowest energy transition has a probability of 0.9 (Liu et al., 1998) and the rest have 0.1. The average collisional probability of the hot atoms with the ambient atmosphere is $P = n(z)\sigma(v)V_{rel}$, where $\sigma(v)$ is the velocity-dependent momentum transfer cross-section, $V_{rel}$ is the relative collision velocity and $n(z)$ is the local H and $H_2$ density at altitude z. $\sigma(v)$ is ~$4\times10^{-16}$ cm$^2$ for $H_2 + H^*$ collisions for $H^*$ with energy of ~4 eV and above (Phelps, 1990). For $H + H^*$ collisions, $\sigma(v)$ is ~$1\times10^{-15}$ cm$^2$ for $H^*$ with energy of a few eV and below. In a Monte-Carlo simulation, at each time step a random number R is chosen to decide if a collision occurs. If one collision occurs (which means R < P), the energy loss from the hot H due to hard-sphere collision with the H and $H_2$ particles is $\Delta E_k = R' \frac{4 m_1 m_2}{(m_1 + m_2)^2} E_k$. The value R' is a random number between 0 and 1, and $m_1$ and $m_2$ are the masses of the colliding particles.

At the exiting altitude (the exobase), the kinetic energy distributions of hot H after collisions with the ambient H and $H_2$ at different depths below exobase are shown in Figure. 3b. For the distribution at 1.5 scale heights below exobase, it is seen that the primary peak is at ~0.1 eV plus a small peak near 3 eV. In addition, the lower the source location of hot H, the more energy is lost before it reaches the exobase and the less likely escapes. Here we simply use one averaged distribution of these three different H-forming locations (0.5, 1 and 1.5 scale heights below the exobase). Therefore, each test particle in our computation of ballistic trajectories of the hot H from Saturn's exosphere is assigned an initial ejection velocity from the energy distribution derived above using with a Monte Carlo scheme. This is in addition to the local corotation velocity consistent with the latitudinal source point taking account of Saturn's oblateness, which was found to be very important (Shemansky et al. 2009). Both isotropic emission and radial outflow emission are examined using our simulations.

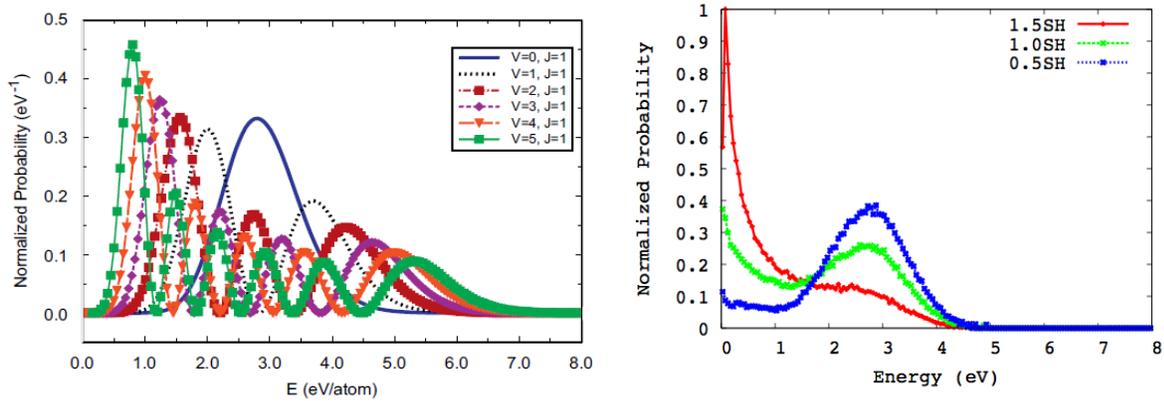

Figure 3: a(left): Probability distributions of H kinetic energy for electron impact at asymptotic energy on $H_2$ X state. It is adopted from Figure 17 in Shemansky et al. (2009). b(right): The kinetic energy distribution of hot H after collisional thermalization with the ambient H and $H_2$ for three different source altitudes for hot H: 0.5 (red), 1.0 (green) and 1.5(blue) scale heights below the exobase.

**2.5.2 Initial Launch Position on Saturn**

The H source region is assumed to be in the Saturn's sunlit hemisphere (Shemansky et al., 2009). To compare with Cassini UVIS data, we use the geometrical condition at SOI for which the subsolar point was at ~ -24° latitude and the phase angle between Sun and Cassini was ~77°. However, the putative plume structure for that configuration was discovered to peak at ~-13.5° latitude. In our simulations, the test particle will be launched in the sunlit hemisphere centering in -24° latitude, ~347° longitude (270° is sub-spacecraft longitude) and local time at noon with a Gaussian probability distribution with the FWHM=0.56 $R_S$ (Shemansky et al., 2009). More discussions about initial launch position on Saturn's surface will be presented in next section.

## 3. Results and Discussions

### 3.1 Saturn's H plume

The two modeled distributions of the Saturn's source of H, first as a plume with radial outflow and then as ejections in random directions, are presented in X-Z plane with a color scale of the H Ly-α intensity (R) in Figure 4a and 4b, respectively. X-axis is in the East-West direction (+X in the sunlit side), Z-axis is the North-South spin axis, and Y-axis is along line of sight of the Cassini spacecraft. The simulated column density along line of sight in the same configuration of SOI (sub-spacecraft latitude=0º and subsolar latitude=-24 °) was directly converted to the H Ly-α intensity, Rayleigh (1 R ~5x10$^{10}$ H cm$^{-2}$ from Melin et al., 2009) without considering the orientation of Sun. In spite of this, the H column density can be approximately proportional to the Ly-α intensity (R), except some small, specific regions such as in Saturn's shadow and the ring shadow region (above the ring plane), with a reasonable assumption that the H cloud is optically thin.

It is clearly seen in Figure 4a that H ejecting from Saturn only in radial outflow directions has a plume-like structure. Otherwise, the morphology becomes a very broad distribution in a spherical shape if the atomic H flows out of Saturn in random directions (seen in Figure 4b). As stated in Shemansky et al. (2009), the physical mechanism of radial outflow is still unclear. It is also found that most of the hot H escaping from Saturn both in radial and random directions would not go out further ~5-10 Rs due to collisional thermalization with the ambient $H_2$ and H in Saturn's atmosphere. A very small fraction (<1%) can travel into the outer magnetosphere. However, the Cassini UVIS data shown in Melin et al. (2009) indicated that the H Ly-α intensity at equator (Z=0) near Titan's orbit was still ~30-50% of the value near Saturn. This comparison suggests that Saturn is not the only source of the atomic H cloud in the Saturnian magnetosphere.

Figure 4f shows the H intensity map observed by Cassini UVIS at SOI (which is adopted from Figure 5 in Melin et al., 2009). We note that some features shown on Saturn's surface were actually due to the auroral and dayglow processes (Shemansky et al., 2009). In these comparisons shown below, we ignore the gravitationally bound H at the top of Saturn's atmosphere, as indicated by the Lyman-alpha glow on the illuminated quadrant in Figure 4f. Compared to the UVIS H intensity map, our simulated H plume for radial outflow (Figure 4a) does not show a component in the anti-solar side (-X direction). This might be accounted for if we assume that the plume source region is peaked in different local time. This will be discussed in next paragraph.

In fact, the Saturn atmospheric source region (e.g. size distribution and peak locations such as latitude and local time) is difficult to determine since it is strongly dependent on the ionospheric electron (or photoelectron) density. Many factors are found to influence the electron density distribution in Saturn's ionosphere such as the ring shadow and the dawn/dusk asymmetry. That the plasma density is higher near dusk than near dawn (even at altitudes near the exobase) has been observed by the Cassini Radio Science Subsystem (RSS) (Nagy et al., 2006; Kliore et al., 2009). This dawn-dusk asymmetry is suggested to be related to those mechanisms that convert the long-lived $H^+$ into short-lived molecular ions ($H_3^+$ from McElroy, 1973; $H_2O^+$ from Connerney and Waite, 1984). Since these molecular ions have a relatively short recombination timescale on Saturn's nightside, the ionosphere plasma density is reduced in the dawn sector. Because variations of the ionospheric electron density in local time, the hot H source rate may have its maximum near dusk rather than at noon as initially assumed in the cases of Figure 4a and 4b. The hot H source region centered at different local time can definitely modify the morphology of the hydrogen cloud in the Saturnian magnetosphere. A test case is shown in Figure 4c in which the H plume source region is centered at LT=17 hr with the same Gaussian distribution as described above. The morphology of the resulting H cloud is seen to have a contribution on the anti-solar side (-X direction), which seems more similar to the UVIS map in panel f. In Figure 4d, we show another test case: the same H plume source region assumed in Figure 4c but with the H emission energy assumed to be 4.0 eV (even higher than Shemansky et al., 2009) without collisions between the hot H and the ambient atmosphere. There, a much broader H cloud is seen. It supports that the collisional thermalization of hot H with the ambient atmosphere plays an important role in modifying the H emission velocity distribution out of Saturn and, consequently, the morphology of Saturn's H cloud.

The morphology of Saturn's H cloud obtained using a subsolar source in Figure 4a, 4c and 4 d is peaked at latitude ~ -24° (subsolar point) which does not agree with Cassini UVIS observations at SOI (peak at ~ -13.5°). If our modeling is correct, this suggests that the ionospheric electron density might not have its maximum at subsolar point. With Cassini RSS observations, Moore et al. (2010) showed that the peak electron density and the total electron content exhibit a clear increase with latitude with a minimum at Saturn's equator. They found that the latitudinal variations of Saturn's ionospheric electron density can be explained by an introduction of water flux near equator and particle precipitation (e.g. likely aurora) at high latitudes based on a comparison with the Saturn-Thermosphere-Ionosphere-Model (STIM) simulations, however, most of the high-latitude RSS data are from the region outside of Saturn's typical main auroral oval. In addition, the ring shadow is found to cause strong plasma depletions in the Saturn's ionosphere (e.g. Mendillo et al., 2005; Moore et al., 2010). At SOI, Saturn was in southern solstice so there was a large area of northern hemisphere in the ring shadow. Therefore, at SOI, the ionospheric electron density in the northern hemisphere would be smaller than in the southern. By the same token, the ring shadow effect can enhance the asymmetry on the hot H source rate. Therefore, we would expect that there are seasonal variations in the morphology of Saturn's H ejection, so that one should see an enhanced distribution in the northern hemisphere at northern solstice. This can be verified by the on-going Cassini observations at equinox and at northern solstice (Proximal orbits).

The location and extent of the H source might also account for the fact that the simulated H morphology is broader than that indicted by the UVIS map (e.g. Figure 4a). As described in Sect. 2.5.2, we used a symmetric Gaussian distribution with FWHM=0.56 $R_S$ for the size of the H plume source which seems to be too large. This value was inferred from the line of sight column density observed by UVIS whereas the actual H distribution in 3D could be smeared due to projection on the 2D map. The actual size distribution could be derived from the UVIS H intensity map in 2D with a deprojection method (e.g., Willingale et al., 1996). However, this is beyond scope of our present work. In addition, as mentioned above, the ring shadow can cause strong plasma depletion in the northern hemisphere at SOI so that the H plume source region is likely not symmetric around latitude -24º (or -13.5º).

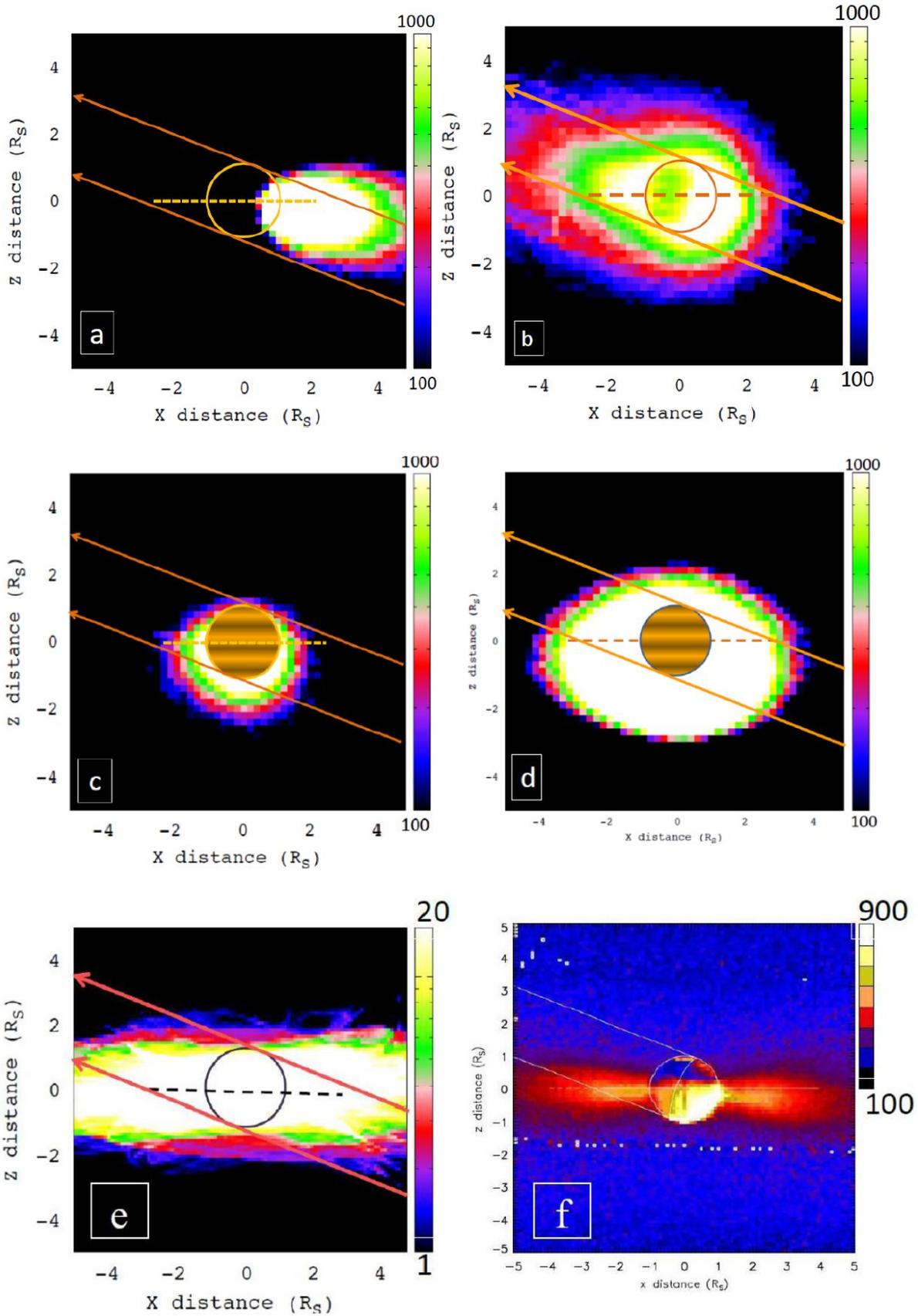
484

Figure 4: Simulated morphologies of the spatial distribution of H produced by escape from Saturn's atmosphere. Presented in X-Z plane with a color scale of the H Ly-α intensity along line of sight in the same configuration at SOI (sub-solar point at ~-24º in the +X direction). X-axis is in the East-West direction, Z-axis is the north-south spin axis, and Y-axis is along line of sight of the spacecraft. Colors represent the H Ly- α intensity (R) *but with different scales*.
(a): H emission in radial outflow and the plume source region peaks at noon.
(b): H emission in random directions and the plume source region peaks at noon. (c): H emission in radial outflow and the plume source region peaks at LT=17 hr.
(d): H emission in radial outflow and the plume source region peaks at LT=17 hr, but with the H emission energy of 4.0 eV without consideration of collisions between the hot H and the ambient atmosphere.
(e): The H cloud from the $H_2$ atmosphere of main rings.
(f): The observed H Ly-α intensity of Saturn's H plume by Cassini UVIS at SOI (which is adopted from Figure 5 in Melin et al., 2009).

### 3.2 The $H_2$ atmosphere of main rings

The morphology of the H cloud from photodissociation of the ring $H_2$ atmosphere at SOI is shown in Figure 4e. Ignoring the very different intensity scale and noting the somewhat different spatial scale, the simulated cloud differs from the Cassini UVIS cloud shown in Figure 4f in some way: a symmetric distribution above and below the ring plane with a peak intensity around equator. Accounting for the illumination from below, the simulated H Ly-α intensity in Figure 4e above the ring plane would be slightly lower because of ring absorption. However, the peak intensity in the Cassini observation at X=+1.9 $R_S$ (on the sunlit side) is ~18,000 km below the ring plane (Melin et al., 2009) which can not be duplicated using our modeled H cloud from dissociation of the ring $H_2$ atmosphere. In addition, the Cassini observation showed a more extended distribution in North-South direction (Z-axis) with the FWHM=1.0 $R_S$ at X=+2.0 $R_S$ (Melin et al., 2009) than our modeling results with the FWHM=~0.5 $R_S$.

It is found that ~30% of H from the ring $H_2$ atmosphere impacted Saturn's atmosphere, ~65% lost due to ring absorption, and ~5% ended by photoionization or escaped from Saturn. Because the H lifetime over the main rings is greatly reduced due to ring absorption, the total number of H inside 5 $R_S$ produced by photodissociation of the ring $H_2$ atmosphere at SOI is only ~5-10% of the number observed by Cassini UVIS. In addition, the H cloud from the ring $H_2$ atmosphere is enhanced in the C-ring region due to less ring absorption. This can be clearly seen in Figure 5 which shows the H Ly-α intensity in an E-W slice at equator (from the case of Figure 4e) with double peaks in the C-ring region (pink line).

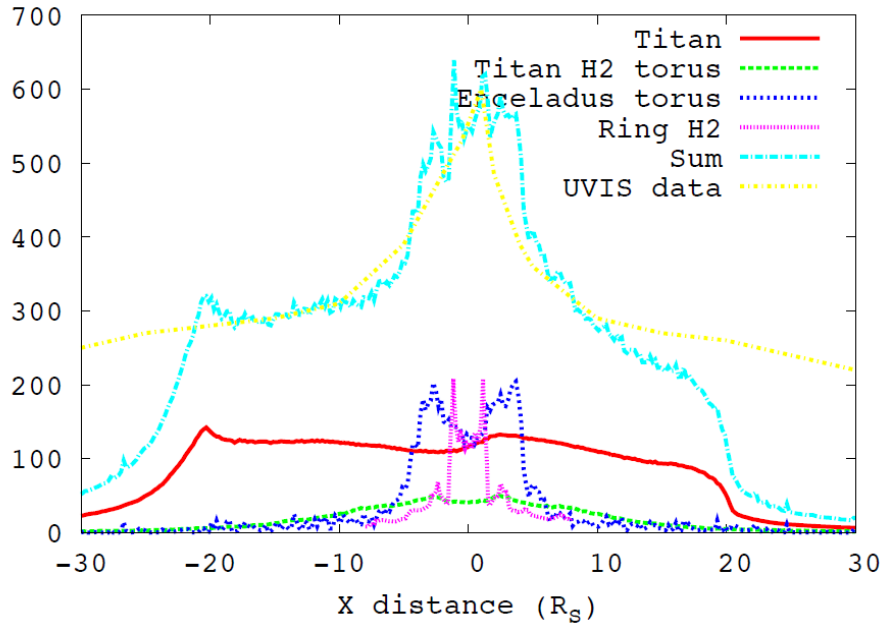

Figure 5: The H Ly-α intensity (R) in an East-West slice (X direction) at equator (Z=0) in the same configuration of SOI: Titan's atomic H torus (red), the H cloud from Titan's $H_2$ torus (green), the H cloud from Enceladus' neutral $H_2O$ and OH torus (blue), the H cloud from the ring $H_2$ atmosphere (pink) and the Cassini UVIS data from Figure 9 in Melin et al. (2009) (yellow). "Sum" (light blue) is the sum of the contributions from the Enceladus' torus, the ring $H_2$ atmosphere, Titan's $H_2$ cloud and a factor of 2.2 times the Titan H torus contribution.

**3.3 Enceladus' $H_2O$ and OH Torus**

The H cloud produced from Enceladus' $H_2O$ and OH torus is a roughly symmetric torus with an enhanced density in inner region. Although the excess energy from photodissociation of $H_2O$ and OH is large, it is still found to be a significant contribution to the atomic H cloud inside ~5 $R_S$ (also seen the blue-dotted line in Figure 5). However, the electron-impact dissociation of Enceladus $H_2O$ and OH torus gives only a small contribution for two reasons. First, the H source rate of electron-impact dissociation is only ~17% of photodissociation (see Table 1). Second, the H production rate by electron impacts peaks around 8-10 $R_S$ (due to hot electrons as mentioned before) a region from which the resulting H can escape from Saturn, while photodissociation mainly produces the fresh H around Enceladus' orbit. The total number of H from Enceladus $H_2O$ and OH torus inside ~5 Rs is about 20% of the number observed by Cassini UVIS.

### 3.4 Titan's $H_2$ torus

Figure 6a shows the column density along the Z-axis of the H cloud produced from Titan's $H_2$ torus on the equatorial plane. The H cloud is roughly azimuthally symmetric. This is because the morphology is primarily determined by electron-impact dissociation, which is about three times the photodissociation source rate. Furthermore, the electron-impact dissociation rate peaks around 8-12 $R_S$ (due to hot electrons) and the H lifetime inside ~13 $R_S$ is smaller than the timescale for the cumulative effect of solar radiation pressure, as mentioned before. Both resulted in the roughly symmetrical distribution. The H Ly-α intensity (R) in an East-West slice (X direction) at equator is shown in Figure 5 (green line). It makes a very minor contribution since the total H source rate from Titan's $H_2$ torus is small (~17% of the source rate from Enceladus' neutral torus; see Table 1).

### 3.5 H directly escape from Titan's exosphere

Figure 6b shows the column density along the Z-direction of Titan's atomic H torus on the equatorial plane. It is also shown a very broad distribution with a local time asymmetry – an enhancement in the predawn side but with a cavity in the inner region ($< 10R_S$). As discussed in previous work (Marconi and Smith, 1993; Ip, 1996), this local-time asymmetry is a result of the cumulative effect of solar radiation pressure turning the initial circular orbit into the highly eccentric orbit over a long period. There is no obvious seasonal variation in the morphology of Titan's H torus since the H lifetime is very long in the outer magnetosphere (~1/4 of Saturn's orbital period). In addition, the H lifetime and the H emission velocity from Titan are important constraints on the morphology and density distribution of Titan's H torus (Ip, 1996). For example, the higher is the H emission velocity escaping from Titan, the smaller is the density of Titan's H torus.

The H Ly-α intensity of Titan's atomic H torus in an E-W slice at equator is shown in Figure 5 (red line). Obviously, it is a leading contributor to the atomic H in Saturn's magnetosphere outside ~5 Rs. It also exhibits an asymmetry in the outer magnetosphere near Titan's orbit – slightly reduced in the sunlit side (+X direction) and a little enhanced in the anti-sunlit side (-X direction). This is because of the line of sight viewing geometry, which is shown in Figure 6b. This kind of asymmetric distribution is also seen in the Cassini UVIS data at large distance on the –X side (Melin et al., 2009) (yellow line in Figure 5). Assuming these features are connected suggests that the H escape rate of Titan (~$1.74 \times 10^{27}$ s$^{-1}$) used in our simulation is underestimated. Therefore, in obtaining the "Sum" in Fig. 5 we multiply this contribution by a factor of 2.2 to normalize to the Cassini UVIS data.

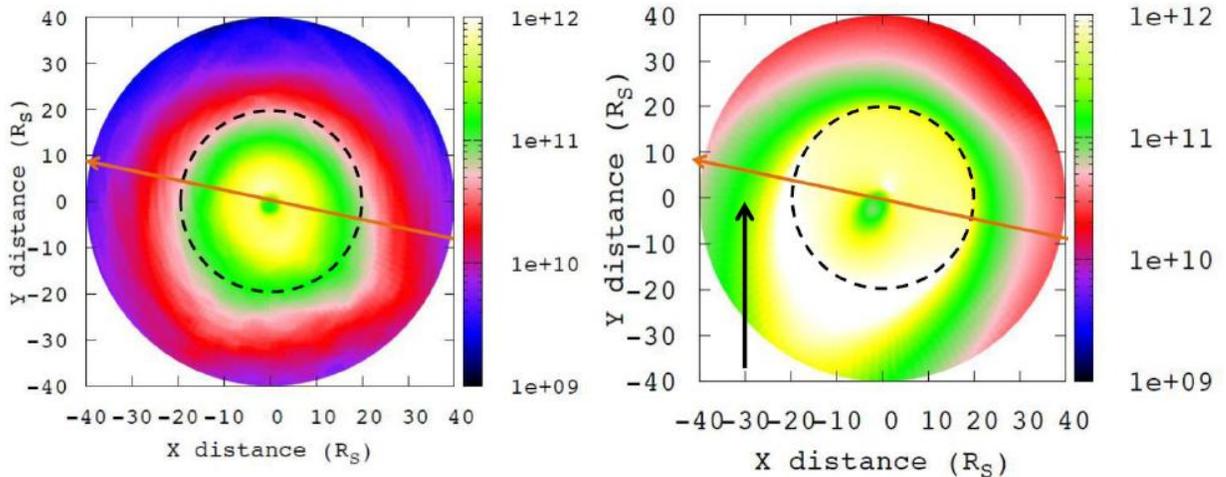

Figure 6: a (left): the column density (#/cm$^2$) along the Z-direction of the H cloud produced from Titan's H$_2$ cloud on the equatorial plane. b (right): the column density (#/cm$^2$) along the Z-direction of Titan's atomic H torus on the equatorial plane. Dashed Circle is Titan's orbit. Orange arrow is Noon-Midnight line (Sun is on the right). Black arrow is the line of sight of Cassini at SOI.

### 3.6 Overview

Titan's H torus is shown to be the major source of the H cloud in Saturn's outer magnetosphere. In inner region (<5-10 R$_S$), Saturn's exosphere, the H$_2$ atmosphere of main rings and Enceladus' H$_2$O and OH torus are contributors as well. It is seen, in Figure 5, that a sum of the contributions of the H cloud from the ring $H_2$ atmosphere, Enceladus' neutral torus, Titan's $H_2$ cloud and directly from Titan which is multiplied by a factor of 2.2 (light-blue line) seems roughly consistent with the Cassini UVIS data (yellow line) between 10-20 $R_S$ with a better agreement in the anti-solar side.

On the sun-lit side, our results indicate a slight excess inside 10 $R_S$ and a deficiency at 10-20 $R_S$ which might be filled by the high-velocity (or escaping) component of Saturn's H atoms. Alternatively, some mechanisms involving velocity re-distribution can change the morphology of the H cloud. Neutral-neutral collisions around Enceladus' orbit have been found to significantly change the density distribution of its neutral $H_2O$ and OH torus (Farmer, 2009; Cassidy and Johnson, 2010). Similarly, the morphology of the H cloud from Enceladus' torus will be modified by the collisions of H atoms with neutral $H_2O$ and OH, which is not included in the current simulations. But, this would not make a big impact on the density distribution of Titan's H torus since the H escaping from Titan spends less time in inner region. However, it might account for the differences in the inner region (<10 $R_S$) and needs further investigation.

Outside Titan's orbit, the simulated intensity is lower than the UVIS data (seen in Figure 5). One possible explanation is that the H ejection velocity from Titan is higher than that produced by thermal escape from the exosphere due to the large excess energy from photodissociation of $CH_4$ and $H_2$. . Alternatively, this difference could be a due to interplanetary H or the interaction between the H and solar wind ions which is neglected in the present simulation. To be consistent with the UVIS data (inside ~ 20 $R_S$), Titan's H escape rate near SOI is suggested to be $3.8 \times 10^{27}$ H $s^{-1}$ at T=150K which agrees with the estimate (3.3 - 4.8 x $10^{27}$ H $s^{-1}$ at T=186K) in Smyth and Marconi (2006) using with the Voyager data. In order to fit the H observations, they also suggested a source rate of 1.4 - 1.9 x $10^{28}$ H $s^{-1}$ from the inner icy satellite region. Our estimate showed the H source rate from Enceladus neutral $H_2O$ and OH torus is $\sim 7 \times 10^{27}$ $s^{-1}$ with a contribution from Saturn's atmosphere and the ring $H_2$ atmosphere.

Because of the large uncertainty on the morphology of the H source from Saturn's atmosphere (e.g. size distribution, peak locations of local time and of latitude, and the H emission velocity distribution) it is difficult to draw a firm conclusion about Saturn's contribution to the UVIS observations. In general, the H flowing out of Saturn is found to populate the H cloud only inside ~5-10 $R_S$. The results in Figure 4 indicate that the hydrogen density from Saturn's atmosphere is enhanced on the sunlit side, but it clearly cannot account for the enhancement in the anti-solar side near Titan's orbit seen in the Cassini UVIS data. Therefore, Saturn's atmosphere is not the only uncertain source for the H cloud. In a summary, our estimate of the contributions to the total number of H in the region between 1-5 $R_S$ (not including the component on Saturn's disc seen by UVIS) are: ~5-10% from ring $H_2$ atmosphere, ~20% from Enceladus neutral torus, ~50% from Titan's H torus and implying ~20% from Saturn's atmosphere. Since H from the $H_2$ ring atmosphere has a peak intensity at the equator, for the sources examined only H flowing out of Saturn's atmosphere can account for the peak intensity occurring well below the ring plane. This is the case unless there is a significant direct H source from the ring particles as discussed below.

### 3.7 Other Possible Sources

The discovery of the H Ly-α emission in vicinity of Saturn's main rings occurred about four decades ago (Weiser et al., 1977). It was suggested earlier that the H cloud was produced directly from the ring particles by sputtering by radiation-belt protons (Cheng and Lanzerotti, 1978), by photosputtering (Carlson, 1980), or by the neutralization of Saturn's ionospheric $H^+$ outflow on the ring surface (Ip, 1978). Unfortunately, the interactions between $H/H^+$ and the icy ring particles (e.g. the efficiency of absorption and of neutralization) is not well understood yet. Although H is produced by photo-desorption from ice, as proposed by Carlson (1980), the suggested H source rates from each of the above processes appear to be too low to account for the observed abundance. This is due to the very short lifetime for H which is typically assumed to be lost to ring absorption in one ballistic transport. We estimate that a source ~$10^{29}$ H $s^{-1}$ directly from the rings would be needed to account for the Cassini UVIS feature below the ring plane near SOI (several hundred Raleigh). As discussed in our previous work (e.g. Tseng et al., 2012), Saturn's magnetosphere was extremely compressed at SOI and there was strong solar activity in 2004. As a result, some transient or sporadic event which enhanced ring particle decomposition could have released unusually large amount of H from the rings which might account for the Cassini observation at SOI. However, such a source of H would peak around the ring plane, which does not appear to agree with the Cassini observations.

Jupiter is also surrounded by hydrogen bulge (Sandel et al., 1980). Because the Jovian hydrogen bulge co-rotated with the magnetic field, an internal or upward flow of H was not suggested. Rather was proposed to be formed by magnetospheric electron bombardment on Jupiter's atmosphere (Sandel., 1980; Clarke et al., 1980). In addition, the hydrogen bulge with longitudinal and latitudinal asymmetry (enhanced in the 8-12° above the equator) was suggested to be a direct evidence of co-rotating magnetospheric convection pattern with the magnetic-anomaly effect in Jupiter's northern hemisphere (Dessler et al., 1981). This would not be applicable to Saturn's H cloud since no magnetic anomaly has been found in Saturn's magnetic field, and, currently, there is no observational data showing that Saturn's plume co-rotates with the magnetic field.

## 4. Summary

In this work we have carried out a comprehensive investigation of the H cloud in Saturn's magnetosphere taking into account a number of sources: Saturn's exosphere, the $H_2$ atmosphere of main rings, Enceladus' OH and $H_2O$ torus, Titan's $H_2$ torus and H directly escaping from Titan as summarized below.

**4.1 Saturn's exosphere:** The H source region (e.g. size distribution and peak location such as local time and latitude) on Saturn's disk is an important constraint on the morphology of Saturn's H cloud. As shown in our simulations (e.g. Figure 4), the H source region would be centered in the late afternoon area (local time) in which the morphology of Saturn's H cloud is similar to the Cassini UVIS observation. It is also consistent with the electron density distribution in Saturn's ionosphere with an enhanced density on the dusk side if the atomic H is produced by electron-impact dissociation of $H_2$. We also showed that the collisions between hot H and the ambient atmospheric H and $H_2$ of Saturn significantly modify the H ejection velocity and angle distribution flowing out of Saturn, and, hence, the morphology of Saturn's H cloud as well. However, the H must be emitted as a radial outflow to sustain the putative plume-structure seen by UVIS. Unfortunately, this physical mechanism is still unclear as stated in Shemansky et al. (2009). Finally, it is still hard to draw a conclusion about Saturn's contribution since the H source region is not well defined which severely affects the morphology of Saturn's H cloud.

### 4.2 Other sources:

**1. The $H_2$ ring atmosphere**: Although the simulated morphology of the H cloud (seen in Figure 4e) is shown to be roughly similar to the Cassini UVIS map, the H intensity is much too small since it is quickly lost to ring absorption. To account for the UVIS observations, an H source rate of the order of magnitude of $10^{29}$ $s^{-1}$ directly from the rings is required. An extremely large transient could result from enhanced photolytic decomposition under extreme conditions as discussed above. However, such a source would peak near the ring plane which is not seen in the observations. Therefore, an *enhanced* H source directed downward form the sunlit side, like the photo-desorption of H suggested by Carlson (1980), is a possibility.

**2. The Enceladus and Titan Tori:** The Enceladus $H_2O$ and OH torus is an important contributor inside 5 $R_S$, but the H from Titan's $H_2$ torus is not.

**3. Titan's atomic H torus**: It is a leading contributor in the outer magnetosphere as well as an important contributor in the inner region as discussed below. Its morphology, shaped by solar radiation pressure and Saturn's oblateness, showed a local time asymmetry with an enhancement in predawn side. In addition, because of the viewing geometry of Cassini at SOI, it could explain the asymmetry of the H Ly-α intensity near Titan's orbit observed by Cassini UVIS. However, to account for the Cassini data, Titan's H escape rate must be ~$3.8 \times 10^{27}$ $s^{-1}$ if flux is a Maxwellian at T=150K, which agrees well with the estimate by Smyth and Marconi (2006).

**4.3 Summary Inside 5 $R_S$:** The total number of H observed by UVIS inside 5 $R_S$ is $5.7 \times 10^{34}$ (Shemasnky et al., 2009). Although neutral-neutral collisions in the Enceladus neutral torus will modify the morphology of the H cloud in inner region, the present work shows that ~20% is from Enceladus' $H_2O$ and OH torus, ~5-10% is from the $H_2$ atmosphere of main rings, ~50% is from Titan's atomic H torus, implying ~20% from Saturn's exosphere which makes the scaling factor of the H source rate of Saturn, *c*, equal to 0.2.

## Acknowledgement

This work was support by a NASA JPL contract to SwRI for support of the CAPS instrument on Cassini and from a grant from the NASA's Planetary Atmospheres Program.